\documentstyle[aps,prl,multicol,floats,epsfig]{revtex}

\newcommand{\bi}{\begin{itemize}}
\newcommand{\ei}{\end{itemize}}

\newcommand{\be}{\begin{equation}}
\newcommand{\ee}{\end{equation}}
\newcommand{\bestar}{\[}
\newcommand{\eestar}{\]}
\newcommand{\bea}{\begin{eqnarray}}
\newcommand{\eea}{\end{eqnarray}}
\newcommand{\beastar}{\begin{eqnarray*}}
\newcommand{\eeastar}{\end{eqnarray*}}

\newcommand{\lav}{\left\langle}
\newcommand{\rav}{\right\rangle}

\newcommand{\half}{\frac{1}{2}}

\newcommand{\eq}[1]{~(\ref{#1})}

\newcommand{\order}{{{\cal O}}}

\newcommand{\ie}{{\it i.e.}}
\newcommand{\eg}{{\it e.g.}}

\newcommand{\eps}{\epsilon}
\newcommand{\teq}{t_{\rm eq}}
\newcommand{\eql}{_{\rm eq}}

\newcommand{\rate}{\Gamma}

\newcommand{\dbar}{\bar{d}}
\newcommand{\xbar}{\bar{x}}
\newcommand{\Tin}{T_{\rm i}}

\newcommand{\p}{P}

\newcommand{\ddtau}{\partial_\ta}
\newcommand{\ta}{\tau}
\newcommand{\GG}{G}
\newcommand{\Gact}{H}
\newcommand{\height}{h}
\newcommand{\logtime}{\nu}
\newcommand{\ptilde}{\tilde{P}}

\draft

\begin{document}

\author{Peter Sollich$^1$\cite{fellow_email} and Martin R Evans$^2$}

\address{$^1$Department of Mathematics, King's College London, Strand,
London WC2R 2LS, U.K.\\
$^2$Department of Physics and Astronomy, University of Edinburgh,
Edinburgh EH9 3JZ, U.K.}

\title{Glassy timescale divergence and anomalous coarsening
in a kinetically constrained spin chain}

\maketitle

\begin{abstract}

We analyse the out of equilibrium behavior
of an Ising spin chain with an asymmetric kinetic
constraint after a quench to a low
temperature $T$. In the limit $T\to 0$, we provide an exact solution of the
resulting coarsening process. The equilibration time exhibits a
`glassy' divergence $\teq=\exp(\mbox{const}/T^2)$ (popular as an
alternative to the Vogel-Fulcher law), while the average domain length
grows with a temperature dependent exponent, $\dbar \sim t^{T\ln
2}$. We show that the equilibration time $\teq$ also sets the
timescale for the linear response of the system at low temperatures.

\end{abstract}

\vspace{-0.9\baselineskip}

\begin{multicols}{2}
\noindent
Even after decades of research, understanding the dynamics of glasses
remains a challenging problem (see
\eg~\cite{Fredrickson88,Angell95,BouCugKurMez98}). One of
the main features of glassy systems is that their relaxation time
$\tau$ increases quickly as the temperature $T$ is lowered. A popular
representation of this increase (for so-called `fragile' glasses
\cite{Angell95}) is
the Vogel-Fulcher (VF) law,
$\tau\sim\exp[-\mbox{const}/(T-T_0)]$. This predicts that $\tau$
diverges at temperature $T_0$, and the latter has therefore been
associated with the temperature at which a true thermodynamic glass
transition (achievable only in the limit of infinitely slow cooling)
would take place. However, other functional forms for $\tau(T)$ that
have been proposed do not exhibit singularities at any finite $T$,
indicating the absence of a thermodynamic glass transition. Among
these, the
exponential inverse temperature squared (EITS) form
$\tau\sim\exp(\mbox{const}/T^2)$ is popular.  Experimentally, it is
difficult to distinguish between VF and EITS behavior due to obvious
limitations on the longest accessible timescales; both can represent the
experimentally observed $\tau(T)$ in many
materials~\cite{RicBaes90}. Thus 
analytical results are desirable to shed light on this controversy.
In this work we solve a simple dynamical model exhibiting glassy
dynamics and find EITS behavior.

To model relaxation in glassy systems theoretically, one can postulate
some kind of {\it quenched disorder}, either in terms of some underlying
microscopic Hamiltonian (as is done in spin glasses) or more
phenomenologically by making assumptions about the phase space of the
system (\eg\ in terms of hierarchical or ultrametric
structures~\cite{PalSteAbrAnd84,BouDea95} or energy barrier
distributions~\cite{Vilgis90,BouComMon95}). 
So far the main theoretical justification for either VF or EITS
behavior comes from the latter approach; the EITS law, for example,
is motivated by considering activated dynamics in a landscape of
Gaussian distributed energy barriers~\cite{Vilgis90}.

The alternative approach is to
consider simple 
models whose {\em dynamics} directly induce glassiness.
Examples include systems with kinetic
constraints~\cite{Fredrickson88} or
entropic barriers~\cite{Ritort}, and driven diffusive 
models~\cite{EKKM}. Such an approach is more
obviously relevant to the dynamics of structural glasses (where
quenched disorder is absent) since one does
not need additional arguments that relate quenched and dynamically
`self-induced' disorder~\cite{BouMez94}. The present work
provides a first example where EITS behavior emerges directly
from a {\em microscopic} model {\em without} imposed quenched disorder;
instead energy barriers arise naturally
from dynamical constraints.

We consider a chain of spins in a uniform field,
whose dynamics is nontrivial due to an asymmetric kinetic constraint.
This model was
introduced by J{\"a}ckle and Eisinger~\cite{JaecEis91} and has
recently been rediscovered~\cite{MunGabInaPie98}. We study in
particular the behavior after a quench to a low temperature $T\to
0$. We solve the resulting coarsening dynamics exactly in this limit
and find two main results: Firstly, 
the equilibration time of the system
diverges as $\teq\sim\exp(1/T^2\ln 2)$ (EITS behaviour).
Secondly,
before equilibrium is reached, the average domain
length grows as $\dbar \sim t^{T\ln 2}$, with an exponent that varies
continuously with temperature.
This novel anomalous coarsening is  a consequence of the dynamical constraint,
which produces scale-dependent energy barriers which grow as the logarithm
of the  domain size.
Finally, we show that $\teq$ is not just the timescale for
equilibration after a quench, but in fact is also the timescale
for relaxation of
spin-spin correlations in equilibrium (at low $T$);
this relaxation
time therefore also has an EITS divergence at low $T$.

The model comprises a chain of
$L$  spins $s_i\in\{0,1\}$
where $1\leq i \leq L$;
periodic boundary conditions imply that the left
neighbor of $s_1$ is $s_L$.
The dynamics for a given temperature $T$ are defined as follows: At
any time, only spins whose left neighbor is up (\ie, has the value 1)
can flip.  For such `mobile' spins, the rate for
down-flips $1\to 0$ is 1, while the rate for up-flips $0\to 1$ is
$\eps=\exp(-1/T)$. Detailed balance is obeyed, and the stationary
distribution is the Boltzmann distribution for the
trivial Hamiltonian
%
$H = \sum_{i=1}^L s_i$.
%
For low temperatures the equilibrium
concentration $c=\eps/(1+\eps)$ of up-spins is small.
Since these spins facilitate the dynamics, the
system evolves slowly for small $T$. Moreover to eliminate an up-spin 
one first has to generate an adjacent up-spin.
\end{multicols}
\twocolumn
\noindent
 Thus there are energy barriers
in the system's evolution.

We will be interested mainly in the behavior after a quench from
equilibrium at some high initial
%
%
temperature $\Tin\stackrel{>}{\sim}1$
to $T\ll 1$.
The basic objects that we use for the
description of the
system are 
{\em domains}. As shown by the vertical
lines in
$
\ldots1|0001|1|1|01|001|1|1|01|0\ldots,
$
a domain consists of an up-spin and all the down-spins that separate it
from the nearest up-spin to the left. The length $d$ of a domain also
gives the distance between the up-spin at its right edge and the
nearest up-spin to the left. Note that adjacent up-spins are counted
as separate domains of length $d=1$. In equilibrium, the distribution
of domain lengths and its average are
\be
\p\eql(d)=\eps/(1+\eps)^d, \quad \dbar\eql=1+1/\eps\; .
\label{pd_equil}
\ee
%
Now consider what happens after a deep quench to $T\ll 1$, $\eps\ll
1$. The equilibrium concentration of up-spins at the final temperature
$T$ is $c=1/\dbar=\eps+\order(\eps^2)$; hence the equilibrium
probability of finding an up-spin within a chain segment of {\em
finite} length $d$ is $\order(d\eps)$ and tends to zero for $\eps\to
0$. In this limit ($\eps\to 0$ at fixed $d$), the flipping down of
up-spins therefore becomes {\em irreversible} to leading order. In
terms of domains, this means that the coarsening dynamics of the
system is one of coalescence of domains: an up-spin that flips down
merges two neighbouring domains 
into one large domain. During such an irreversible coarsening
process, no correlations between the lengths of neighboring domains
can build up if there are none in the initial state~\cite{BDG}.  For
the present model the equilibrated initial state consists of domains
independently distributed according to (\ref{pd_equil}). Therefore a
`bag model' \cite{BDG} or `independent interval approximation' for the
dynamics, which is defined by neglecting correlations between domains,
becomes exact in the low-temperature limit (always taken at fixed
$d$).  

We now estimate the typical rate $\rate(d)$ at which domains of length
$d$ disappear by coalescing with their right neighbors.  Because
domain coalescence corresponds to the flipping down of up-spins,
$\rate(d)$ can also be defined as follows. Consider an open spin chain
of length $d$, with a `clamped' up-spin ($s_0=1$) added on the left.
Starting from the state $(s_0,s_1, \ldots, s_d)$ = $10\ldots 01$,
$\rate^{-1}(d)$ is the typical time needed to reach the empty state
$10\ldots 00$ where spin $s_d$ has `relaxed'. 
Any instance of this relaxation process
can be thought of as a path
connecting the two states. Call the maximum number of `excited' spins
(up-spins except $s_0$) encountered along a path its height $h$. One
might think that the relaxation of spin $s_d$ needs to proceed via the
state 11\ldots 1, giving a path of height $d$. In fact, the minimal
path height $h(d)$ is much lower and given by
\be
\height(d)=n+1 \quad \mbox{for}\ 2^{n-1} < d \leq 2^n
\label{hierarchy}
\ee
where $n=0, 1, \ldots$ This result is easily understood for 
$d=2^n$~\cite{MauJaec99}. To relax the $2^n$-th spin $s_{2^n}$,
one can first flip up $s_{2^{n-1}}$ and use it as an `anchor' for
relaxing $s_{2^n}$. The corresponding path is (with $s_{2^{n-1}}$ and
$s_{2^n}$ underlined)
$1\ldots\underline{0}\ldots\underline{1}$ $\to$
$1\ldots\underline{1}\ldots\underline{1}$ $\to$
$1\ldots\underline{1}\ldots\underline{0}$ $\to$
$1\ldots\underline{0}\ldots\underline{0}$
and reaches height $h(2^n)=h(2^{n-1})+1$; the $+1$ arises because the
anchor stays up while the spin $2^{n-1}$ to its right is
relaxed. Continuing recursively, one arrives at $\height(2^n) =
\height(1)+n$; but $\height(1)=1$ because the only path for the
relaxation of $s_1$ is $11\to 10$. To prove\eq{hierarchy} more
generally, define $d(h)$ as
the length of the largest single domain that can be relaxed 
by a path of height $\leq h$. Because of detailed balance, 
any relaxation path can be reversed, yielding a path of the same
height from the empty state to the state $10 \ldots 01$. 
In the same way, let us define 
$l(h)$ to be the maximal length of {\em any} spin configuration 
(ending in an up spin) that can be reached from the empty state
by a path of height $\leq h$. One then has
$d(h+1)=l(h)+1$ because to relax  $s_{d(h+1)}$
one needs to flip up its left neighbor while exciting no more than $h$
additional spins. A second relation is obtained from the relaxation of
a configuration realizing  the bound $l(h)$.
%
%
Such a configuration contains $h$
excited spins (due to its maximal length).
To relax the first of
these, no extra excitations are allowed (because of the ceiling $h$ on
path height); for the relaxation of the 2$^{\rm nd}$, 3$^{\rm
rd}$\ldots\ $h$-th spin, a maximum of 1, 2\ldots\ $h-1$ excitations
are available. Summing the maximal length change at each step then
gives $l(h)=\sum_{h'=0}^{h-1} d(h'+1)$. The above two recursions for
$l(h)$ and $d(h)$, combined with $d(1)$ $=$ $l(1)$ $=$ 1, yield
$l(h)=2^h-1$ and $d(h)=2^{h-1}$, proving\eq{hierarchy}.

At this stage we already see the key feature of the dynamics:
the energy barrier for the relaxation
of spin $s_d$ is $\height(d)-1$ (the $-1$ comes from the
one excited spin ($s_d$) in the initial state). The
rate for this relaxation is therefore $\rate(d) =
\order(\exp[-(\height(d)-1)/T]) =
\order(\eps^{\height(d)-1})$~\cite{entropy_irrelevant}.
Then eq.~(\ref{hierarchy})  tells us that
the relaxation rate for domains of size $d$
is $\rate(d) \sim   \exp(- \ln d/T \ln 2)$.
Thus the energy barrier for the growth of domains 
increases logarithmically with domain size, giving a 
typical domain size growing
as $\dbar \sim t^{T \ln 2} $. Also, since $\dbar\eql \sim \exp(1/T)$
the equilibration time will grow according to an EITS law
$t_{\rm eq} \sim \exp(1 /T^2 \ln 2)$.

From the scaling of $\rate(d)$, the
coarsening dynamics in the limit $\eps\to 0$ naturally
divides into stages distinguished by $n=\height(d)-1=0, 1,
\ldots$ During stage $n$, the domains with lengths $2^{n-1}<d\leq
2^{n}$ disappear; we call these the `active' domains. This process
takes place on a timescale of
$\order(\rate^{-1}(d))=\order(\eps^{-n})$; because the timescales for
different stages differ by factors of $1/\eps$, we can treat them
separately in the limit $\eps\to 0$. During stage $n$, the
distribution of inactive domains ($d>2^n$) changes only because such
domains can be created when smaller domains coalesce. Combining this
with the (exact) bag model discussed above, we have
for $d> 2^n$
\be
\ddtau \p(d,\ta) = \sum_{ 2^{n-1}< d'\leq 2^n }
\p(d-d',\ta)\,[-\ddtau \p(d',\ta)]\; .
\label{eqn_motion}
\ee
The term in square brackets is the rate at which active domains
disappear; $d'\leq 2^n$ because inactive domains do not disappear.
We use the rescaled
time $\ta=t\eps^n$; during stage $n$ of the dynamics and in
the limit $\eps\!\to\! 0$, it can take on any positive value $\tau>0$.
The initial condition for\eq{eqn_motion} is the domain length
distribution at the end of stage $n-1$ of the dynamics, which we call
$\p_n(d)$ = $\p(d,\ta\!\to\! 0)$. To calculate $\p_{n+1}(d)$ =
$\p(d,\ta\!\to\!\infty)$, introduce the generating function
$\GG(z,\ta)=\sum_{2^{n-1}< d} \p(d,\ta)z^d$, 
and its analog for the active
domains, $\Gact(z,\ta)=\sum_{2^{n-1} < d\leq
2^n}\p(d,\ta)z^d$. From\eq{eqn_motion}, one then finds
\bestar
\ddtau [\GG(z,\ta)-\Gact(z,\ta)] = - \GG(z,\ta)\,\ddtau \Gact(z,\ta).
\eestar
This can be integrated to give $[1-\GG(z,\infty)]/[1-\GG(z,0)] =
\exp[\Gact(z,0)-\Gact(z,\infty)]$. But at the end of stage $n$, all
domains that were active during that stage have disappeared, and so
$\Gact(z,\infty)=0$. Defining the initial condition for $\GG$ as
$\GG_n(z)\equiv \GG(z,0)= \sum_{2^{n-1}<d}\p_n(d)z^d$ and
similarly for the active generating function
$\Gact_n(z)\equiv \Gact(z,0)$,
we then have finally
\be
\GG_{n+1}(z)-1 = [\GG_n(z)-1]\exp[\Gact_n(z)]\;.
\label{main}
\ee
\begin{figure}
%
\epsfig{file=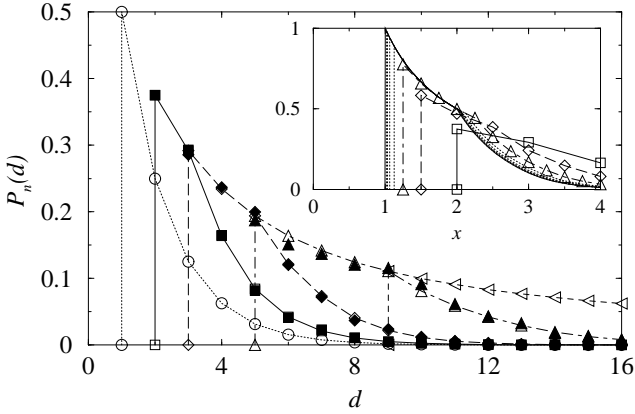, width=8.4cm}
\caption{Domain length distributions $\p_n(d)$ at the end of stage
$n-1$ of the low $T$ coarsening dynamics, for initial temperature
$\Tin=\infty$. Open symbols and lines: Theoretical results,
calculated from~(\protect\ref{main}), for $n=0$ ($\bigcirc$; initial
condition), 1 ($\Box$), 2 ($\Diamond$), 3 ($\triangle$). Full symbols:
Simulation results for a chain of length $L=2^{15}$ and $\eps=10^{-4}$
($n=1, 2$) and $\eps=10^{-3}$ ($n=3$). Inset: Scaled predictions
$2^{n-1}\p_n(d=2^{n-1}x)$ vs.\ $x$ for $n=1, \ldots, 8$. Bold line:
Predicted scaling function~(\protect\ref{scaling_limit}).
\label{fig:pd}
}
\end{figure}
This exact result relates the domain length distributions $\p_n(d)$
and $\p_{n+1}(d)$ at the end of stages $n-1$ and $n$ of the dynamics,
as expressed through their generating functions. Iterating it from a
given initial distribution $\p_0(d)$ gives $\p_n(d)$ for
all $n=1, 2, \ldots$ We do this numerically by expressing\eq{main}
directly in terms of the probability distributions; the exponential is
thus expanded into a series of convolutions of increasing
order. Fig.~\ref{fig:pd} shows the results for the case where
$\p_0(d)$ is the equilibrium distribution\eq{pd_equil} corresponding
to an initial temperature of $\Tin=\infty$. Not unexpectedly, a
scaling limit is approached for large $n$: The rescaled distributions
$\ptilde_n(x)=2^{n-1}\p_n(d)$, where
the scaled domain size is $x=d/2^{n-1}$, converge to a limiting
distribution $\ptilde(x)$ which is independent of the initial
condition.  Invariance under\eq{main} gives an equation
for the corresponding  Laplace transforms
$g(s)$ and $h(s)$ of $\ptilde(x)$
\[
g(2s)-1 = [g(s)-1]\exp[h(s)]\;.
\]
 We 
find a self-consistent solution
%
\bea
\ptilde(x)&=&
\sum_{m=1}^\infty
\frac{(-1)^{m-1}}{m!} \int_1^\infty\! \prod_{r=1}^{m}\frac{dx_r}{x_r}
\ \delta\left(\sum_{s=1}^{m}x_s-x\right)
\label{scaling_limit}
\\
&=&\Theta(x-1)\,\frac{1}{x}-\Theta(x-2)\,\frac{\ln(x-1)}{x}+\ldots
\nonumber
\eea
where $\Theta(x)$ is the Heaviside step function.
This series 
has singularities in the $k$-th derivative
at the integer values $x=k+1$, $k+2$, \ldots
The calculated $\ptilde(x)$ agrees well with the
previous results obtained by direct iteration of\eq{main}
(Fig.~\ref{fig:pd}).
The average
domain length in the scaling limit is given by $\dbar_n=2^{n-1}\xbar$;
from the results for $\ptilde(x)$ we find
$\xbar=\exp(\gamma)=1.78\ldots$, where $\gamma$ is Euler's
constant.

In order to compare the results
to simulations, consider starting from an
equilibrated state at some initial temperature, say $\Tin =\infty$,
quench the system to temperatures $T \ll 1$ at time $t=0$ and
observe its time evolution.
If the results
are plotted against the scaled 
time variable $\logtime=\ln(t)/\ln(1/\eps)=T\ln
t$, then for $T\to 0$ the $n$-th stage of the dynamics shrinks to the
point $\logtime=n$. In this limit we predict that, for
$n-1<\logtime<n$, the domain length distribution is $\p_n(d)$ as
defined by the recursion\eq{main}. The average domain length $\dbar$
will follow a `staircase' function, jumping at
$\logtime=n$ from $\dbar_n=\sum_d \p_n(d)d$ to $\dbar_{n+1}$. In the
large $\logtime$ scaling regime, this tells us that $2^{\logtime-1}\xbar
\leq \dbar \leq 2^{\logtime}\xbar$
(where $\xbar=1.78\ldots$ from above), or $\half\leq \dbar/(\xbar\, t^{T
\ln 2}) \leq 1$ when expressed in terms of ordinary time $t$.

We can therefore say that the system coarsens
with an exponent that depends  on temperature and is
given by $T \ln 2$ to lowest order in $T$. By extrapolating this
coarsening law to the equilibrium domain length
$\dbar\eql=\exp(1/T)+\order(1)$, we then also have that the dominant
divergence of the equilibration time of the system for $T\to 0$ is
$\teq=\exp(1/T^2 \ln 2)$.

\begin{figure}
\vspace*{2.5mm}
\hspace*{1mm}
\epsfig{file=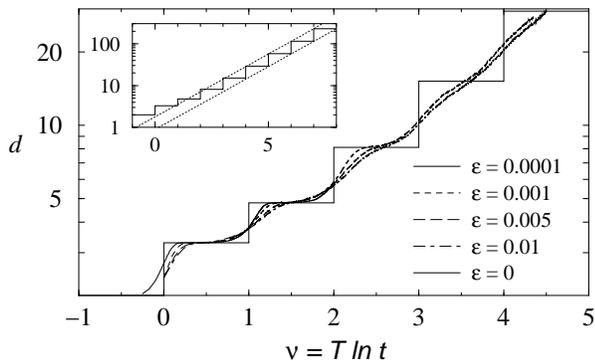, width=7.8cm}

\caption{Evolution of average domain length $\dbar$ after quench from
$\Tin=\infty$ to $T$ at $t=0$, plotted on a log scale vs.\ $\logtime=T
\ln t$.  Simulation results for four values of $\eps=\exp(1/T)$ are
shown, obtained from a single run for a spin chain of length
$L=2^{15}$.  Bold line: Theoretical prediction for $T\to 0$. Inset:
Theory for larger $\logtime$ and $\logtime\to\infty$ asymptotes.
\label{fig:dbar}
}
\end{figure}

In Fig.~\ref{fig:dbar}, we show the results of simulations for a range
of values of $\eps=\exp(1/T)$.  We used a waiting time Monte Carlo
algorithm \cite{BKL}
combined with an
efficient binary tree representation for the positions of the mobile
spins. This let us access far
larger systems ($L=2^{15}$) and longer times
(up to $t=10^{10}$) than in previous
simulations~\cite{MunGabInaPie98}. The plateaus in $\dbar(\logtime)$
that develop with decreasing $\eps$ can clearly be seen, and their
values are in good agreement with the predicted theoretical values. We
also obtained the domain length distributions on the plateaus,
by taking data at the minima of $(d/d\logtime)\dbar(\logtime)$ w.r.t.\
$\logtime$.
These are shown in Fig.~\ref{fig:pd} for the
cases $n=1,2, 3$
%
%
and are again in good agreement with our theory.

Our result for the equilibration time $\teq=\exp(1/T^2\ln 2)$ is based
on the extrapolation of the finite-$\dbar$ coarsening behavior
$\dbar\sim t^{T\ln 2}$ into the equilibrium region
$\dbar=\order(1/\eps)$, where it is no longer strictly valid. We now
show, however, that the same timescale is obtained from the initial
decay of the spin-spin correlation function {\em at equilibrium} at
low temperature $T$. It turns out that due
to the asymmetric constraint the correlation function 
is site diagonal, $\lav
 (s_i(0)-c)(s_j(t)-c) \rav = \delta_{ij} c[R(t)-c]$
~\cite{JaecEis91,EisJaec93}. Here $R(t)$
is the probability that an up-spin at $t=0$ is also up at
a later time $t$. With increasing $t$, it decays from $R(0)=1$ to the
equilibrium concentration of up-spins, $c=\eps/(1+\eps)$. To find the
initial decay of $R(t)$, consider again timescales
$t=\order(\eps^{-\logtime})$ for finite $\logtime$ and $\eps\to
0$. For $\logtime=n+0$, all domains of length $d\leq 2^n$ will have
disappeared.
Therefore only up-spins that bounded longer domains
at $t=0$ will have an
$\order(1)$ probability of still being up.
From the  equilibrium 
distribution\eq{pd_equil}, one sees that they constitute a fraction
$(1+\eps)^{-2^n}$ of the up-spins at $t=0$, and hence
$R(\logtime=n+0) \simeq
1-2^n\eps+\order(\eps^2)$~\cite{Rt_footnote}.
Neglecting 
corrections of $\order(\eps^2)$, the quantity $-\ln R(\logtime)$
thus lies between $2^{\logtime-1}\eps$ and $2^{\logtime}\eps$
(for $\logtime>0$). Reverting to ordinary time, we have
%
$
{\textstyle
1/2\leq -[\ln R(t)]/(t/\teq)^{T\ln2} \leq 1
}
$
%
for short times $(t/\teq)^{T\ln 2}\ll 1$. The relevant timescale that
enters here is exactly the equilibration time $\teq=\exp(1/T^2\ln 2)$
found above. We can thus identify the equilibration time for
coarsening after a quench, with the equilibrium relaxation time; both
have an EITS-divergence at low $T$.

Finally, we discuss briefly the spin-spin
autocorrelation function for longer times $(t/\teq)^{T\ln
2}=\order(1)$, where the analysis becomes more
involved~\cite{JaecEis91,MauJaec99,EisJaec93}. We have tackled this
problem by extending the concept of domains to that of `superdomains'
which are bounded by up-spins that remain up on a given
timescale. Combining this with a plausible hypothesis for the behavior
of the relaxation timescales $\rate^{-1}(d)$ for $d=\order(1/\eps)$,
the following scenario seems likely~\cite{SolEva00}: In the limit $T\to 0$,
$R(t)$ first decays linearly with the rescaled time variable
$\delta=(t/\teq)^{T\ln 2}$. This is compatible to lowest order with a
stretched exponential relaxation. But then the decay becomes much
faster, and $R$ actually decays to zero at a {\em finite} value of
$\delta$. (For nonzero $T$, there is a crossover into a slower decay,
presumably exponential in $t$, at late times.)
It would also be of  interest to study the relaxation times
of similar models in  dimension $D > 1$~\cite{RMJ92}.

{\bf Acknowledgements}: Both authors are grateful for financial
support from the Royal Society.



\begin{thebibliography}{10}

\vspace*{-3.7\baselineskip}

\bibitem[*]{fellow_email}
Email: peter.sollich@kcl.ac.uk.


\bibitem{Fredrickson88}
G.~H. Fredrickson, Ann.\ Rev.\ Phys.\ Chem. {\bf 39},  149  (1988).

\bibitem{Angell95}
C.~A. Angell, Science {\bf 267},  1924  (1995).

\bibitem{BouCugKurMez98}
J.~P. Bouchaud, L.~F. Cugliandolo, J. Kurchan, and M. M{\'{e}}zard,  in {\em
  Spin glasses and random fields}, edited by A.~P. Young (World Scientific,
  Singapore, 1998).

\bibitem{RicBaes90}
R. Richert and H. B{\"{a}}ssler, J.\ Phys.\ Cond.\ Matt. {\bf 2},  2273
  (1990).

\bibitem{PalSteAbrAnd84}
R.~G. Palmer, D.~L. Stein, E. Abrahams, and P.~W. Anderson, Phys.\ Rev.\ Lett.
  {\bf 53},  958  (1984).

\bibitem{BouDea95}
J.~P. Bouchaud and D.~S. Dean, J.\ Phys.\ (France)\ I {\bf 5},  265  (1995).

\bibitem{Vilgis90}
T.~A. Vilgis, J.\ Phys.\ Cond.\ Matt. {\bf 2},  3667  (1990).

\bibitem{BouComMon95}
J.~P. Bouchaud, A. Comtet, and C. Monthus, J.\ Phys.\ (France)\ I {\bf 5},
  1521  (1995).

\bibitem{Ritort}
F.~Ritort,
Phys. Rev. Lett. {\bf 75}, 1190 (1995)

\bibitem{EKKM}
M.~R. Evans, Y. Kafri,  H.~M. Koduvely and D. Mukamel
Phys. Rev. Lett. {\bf 80}, 425 (1998).

\bibitem{BouMez94}
J.~P. Bouchaud and M. M{\'{e}}zard, J.\ Phys.\ (France)\ I {\bf 4},  1109
  (1994).

\bibitem{JaecEis91}
J. J{\"{a}}ckle and S. Eisinger, Z.\ Phys.\ B {\bf 84},  115  (1991).

\bibitem{MunGabInaPie98}
M.~A. Mu{\~{n}}oz, 
A. Gabrielli, H. Inaoka, and L. Pietronero, Phys.\ Rev.\ E {\bf
  57},  4354  (1998).

\bibitem{BDG}
A.~J. Bray, B. Derrida and C. Godr\`eche,
Europhys. Lett. {\bf 27}, 175 (1994);
A.~D. Rutenberg and A.~J. Bray,
Phys. Rev. E {\bf 50}, 1900 (1994).

\bibitem{MauJaec99}
F. Mauch and J. J{\"{a}}ckle, Physica A {\bf 262},  98  (1999).

\bibitem{entropy_irrelevant}
In principle there are also `entropic'
contributions to $\rate(d)$, related to the number of different paths
by which $s_d$ can be relaxed. But as long as we keep $d$ finite when
taking $\eps\to 0$, the number of paths remains finite
and hence does not change the scaling of $\rate(d)$ with $\eps$.


\bibitem{BKL}
A.~B. Bortz, M.~H. Kalos and J.~L. Lebowitz,
J. Comput. Phys. {\bf 17}, 10 (1975)

\bibitem{EisJaec93}
S. Eisinger and J. J{\"{a}}ckle, J.\ Stat.\ Phys. {\bf 73},  643  (1993).

\bibitem{Rt_footnote}
With probability $\order(\eps)$, two short domains of length $d=\order(1)$ will
  be next to each other. The right domain may then not disappear on a timescale
  $\rate^{-1}(d)$ if the left one disappears first and thereby `lengthens' it.
  Also, spins that have flipped down will flip up again with probability
  $\order(\eps)$. We have neglected both of these effects because they only
  give corrections of $\order(\eps^2)$ to $R(\logtime)$.

\bibitem{SolEva00}
P. Sollich and M.~R. Evans (unpublished).

\bibitem{RMJ92}
J. Reiter, F. Mauch and J. J{\"{a}}ckle, Physica A {\bf 184}, 458 (1992)
\end{thebibliography}

\vspace*{-1.0\baselineskip}

\end{document}